\documentclass[11pt,pdftex,dvipsnames,table]{article}
\usepackage[utf8]{inputenc}

\usepackage[top=0.85in,left=0.75in,right=0.75in, footskip=0.75in]{geometry}
\usepackage[utf8]{inputenc}
\usepackage{comment}
\usepackage{blindtext}
\usepackage{hyperref}
\usepackage{amsmath}
\usepackage{amssymb}
\usepackage{graphicx}
\usepackage{subfig}
\usepackage{amsfonts}
\usepackage{caption}
\usepackage{fancyhdr}
\usepackage{float}
\usepackage[none]{hyphenat}
\usepackage[greek,english]{babel}
\usepackage{afterpage}
\usepackage[utf8]{inputenc}
\usepackage[english]{babel}
\usepackage{color}
\usepackage{longtable}
\usepackage{lscape}
\usepackage{lipsum} 
\usepackage{array}
\usepackage{flafter}
\usepackage{color, colortbl}
\usepackage[colorinlistoftodos,prependcaption]{todonotes}
\usepackage[numbers,sort&compress]{natbib}
\usepackage{amsmath,bm}
\usepackage{xargs}                      
\usepackage{svg}
\usepackage{xcolor} 
\usepackage{booktabs,caption}

\title{Meta-analysis of networks of diagnostic tests with binary and continuous results}
\author{Efthymia Derezea$^{1,^*}$, Gabriel Rogers$^{2}$, Nicky J Welton$^{1}$, Hayley E Jones$^{1}$}

\date{April, 2026}

\begin{document}

\maketitle
\begin{center}

	\begin{small}
		$1$. \textit{Population Health Sciences, Bristol Medical School, University of Bristol, UK}\\
		$2$. \textit{Manchester Centre for Health Economics, University of Manchester, UK}\\
        $^*$Corresponding author: e.derezea@bristol.ac.uk
	\end{small}	
	\end{center}
\begin{abstract}
Network meta-analysis of diagnostic test accuracy (NMA-DTA) is a relatively new field, involving combining evidence across studies to evaluate and compare the accuracy of different tests for a given condition. However, the methods proposed to date cannot always capture complex aspects of the data. In fact, many commonly used diagnostic tests are continuous biomarkers, whose accuracy is evaluated at multiple thresholds within a study. Using current NMA-DTA methods we are feasibly able to include in our analysis only a few thresholds per study, discarding this way a big amount of data which could have provided us with useful information. We introduce an approach that can efficiently encompass all available data. This is a hierarchical model that  incorporates multinomial likelihoods for studies reporting results across multiple thresholds and a parametric structure for the relationship between the probability of testing positive and threshold within each disease class. This approach enables us to obtain accuracy estimates of tests across the whole range of observed thresholds, while it retains all the useful properties of standard NMA-DTA methods. We explore different variations of this model based on different covariance structures, the inclusion of study-level random effects, and the addition of a further hierarchical structure on the test-level variance components. This framework is applied to data from two systematic reviews, allowing the inclusion of a larger number of tests (compared to alternative approaches) and estimation of sensitivity and specificity at different thresholds with increased precision.
\end{abstract}

\section{Introduction}\label{Sect:intro}
Diagnostic test accuracy (DTA) studies aim to quantify how accurate a diagnostic test (referred to also as index test) is, for identifying the presence or absence of a particular target condition for an individual \citep{zhou2014statistical}. Usually studies of DTA compare the index test's results against a reference standard assumed to be error-free \citep{trikalinos2012chapter}. Results are summarized in 2x2 contingency tables showing the agreement between the index and the reference test. Based on these, accuracy measures such as the probability of obtaining a positive test result when an individual is diseased (Sensitivity) or disease-free (False positive fraction = 1- Specificity) can be calculated. Increasingly, multiple such studies compare the accuracy of multiple index tests for the same condition, which result in evidence synthesis networks of DTA \citep{walsh2022enamel,holper2024predictive,rogers2025cirrhosis,kawada2024diagnostic} . Methods of network meta-analysis of DTA (NMA-DTA) combine all available evidence for multiple index tests for the same target condition and produce pooled estimates of absolute and relative accuracy \citep{veroniki2022diagnostic}. \par 
NMA-DTA is an emerging field, and unlike network meta-analysis for interventions is much less well established \citep{Macaskill22}. Most of the proposed NMA-DTA methods focus on tests returning binary results that are summarized in 2x2 tables as described above \citep{Takwoingi22,nyaga2018anova,menten2015general,ma2018bayesian,lian2018bayesian}. However, many index tests return results in a continuous spectrum and classification of patients depends on pre-specified cut-off points or thresholds. Many studies focusing on continuous tests report accuracy in multiple such thresholds. For example, in a recent systematic review of tests for detecting hepatocellular carcinoma in patients with cirrhosis \citep{rogers2025cirrhosis}, the alpha-fetoprotein (AFP) test was reported at 157 different thresholds. \par
With current NMA-DTA methods we can efficiently synthesize only a few thresholds per test \citep{owen2018network}. Choosing which thresholds to include in practice is not always straightforward, and most of the time the threshold that is most commonly reported is only included. This approach results in a large amount of evidence being discarded, that could otherwise improve our inference and provide answers regarding the performance of such tests for different thresholds. \par 
In this work, we aim to address this gap in the literature by introducing an NMA-DTA framework for efficiently meta-analyzing both binary and continuous tests. To the best of our knowledge this is the first NMA-DTA model that makes use of all available data of this type, while it enables us answering questions such as how does pooled sensitivity and specificity vary across thresholds, and how do these estimates compare between all tests in the network and for the whole range of observed threshold values.   \par  
The rest of the paper is organized as follows: In Section \ref{Sect:rel_work} we discuss related work for the meta-analysis of a single and networks of index tests. Section \ref{Sect:framework} introduces the hybrid framework for synthesizing both binary and continuous tests, and in Section \ref{Sect:tech} we discuss in more detail technical aspects of these approaches. We apply the model to two real case studies in Section \ref{Sect:applic}. 

\section{Related work}\label{Sect:rel_work}

\subsection{The case of a single test}
When only one index test is available, or of interest, for a target condition, meta-analysis methods for tests returning binary results are well established \citep{Macaskill22}. Analyses are usually based on the Bivariate model \citep{reitsma2005bivariate,chu2006bivariate}, where the true positive (TP) and false positive (FP) counts\footnote{Note that true negative counts could be modeled instead of false positives with both approaches returning identical results.}, $x_{i1}$ and $x_{i0}$ respectively, are assumed to be generated by the following Binomial distributions, independent conditional on the true disease state,
\begin{equation}\label{eq:biv1}
   x_{ij} \sim Binomial(p_{ij}, N_{ij}), \quad j=0,1 \quad i=1,\dots I 
\end{equation}
where $I$ is the total number of studies in a meta-analysis. $p_{ij}$ are the probabilities of a positive test for diseased and healthy and $N_{ij}$ are the number of diseased and healthy individuals in the $i$th study. To account for heterogeneity and between-study correlations among the diseased and healthy, the logit-transformed sensitivities and false positive fractions (FPF) are assumed to follow a Bivariate normal distribution,
\begin{equation}\label{eq:biv2}
  logit(\mathbf{p}_{i}) \sim N(\mathbf{m},\boldsymbol\Sigma)  
\end{equation}
where $\mathbf{p}_{i}=(p_{i1},p_{i2})$, $\mathbf{m}=(m_1, m_2)$ is the mean, and $\boldsymbol\Sigma$ is the $2\times2$ variance-covariance matrix.\par 
A few methods have been proposed for the meta-analysis of tests with continuous results when their accuracy is reported at multiple thresholds per study \citep{jones2019quantifying,steinhauser2016modelling,hoyer2018meta} that rely on quantifying the relationship between accuracy and threshold. In \citet{jones2019quantifying} for example, the TP and FP counts at each threshold $t$, $x_{ijt}$, are assumed to follow a series of conditional Binomial distributions,
$$x_{ij1}|p_{ij1} \sim \mathrm{Binomial}(p_{ij1},N_{ij})$$
\begin{equation}\label{eq:cond_binom}
    x_{ijt}|x_{ij,t-1},p_{ijt}, p_{ij,t-1}\sim \mathrm{Binomial}\left(\frac{p_{ijt}}{p_{ij,t-1}},x_{ij,t-1}\right),\quad t=2,\dots,T_{ij}
\end{equation}
where $T_{ij}$ and $p_{ijt}$ is the total number of reported thresholds, and the probability of a positive test result for threshold $t$, in study $i$ and disease state $j$ respectively. This model further assumes that the latent variable $y_{ijl}$ representing the unobserved test result of patient $l$, transformed by some non-increasing function $g(\cdot)$ follows a logistic distribution, $g(y_{ijl}) \sim Logistic(\mu_{ij},\sigma_{ij})$ which leads to the following relationship between the threshold $C_{it}$ and the probability of a positive test result,
$$logit(p_{ijt})=\dfrac{\mu_{ij}-g(C_{it})}{\sigma_{ij}}$$
To account for heterogeneity, four sets of correlated random effects are placed on the location and scale parameters of latent transformed test results
$$(\mu_{i1},\mu_{i2},log(\sigma_{i1}),log(\sigma_{i2}))^{T} \sim N(\mathbf{m}_{c},\boldsymbol\Sigma_{c})$$
Throughout this paper, we will follow the \citet{jones2019quantifying} framework for modeling continuous tests, but for reasons of mathematical convenience that will become apparent in the following sections, we adopt a slightly different parametrization of the Logistic distribution of the latent test result variables, with a standardized and centered at $C^{*}$ mean, i.e.
$$g(y_{ijl})\sim Logistic(\mu_{ij}*\sigma_{ij}+g(C^{*}),\sigma_{ij})$$
with $C^{*}$ being the most commonly reported threshold. Under this assumption the relationship between a positive test result and threshold is given by
\begin{equation}\label{eq:cent}
 logit(p_{ijt})=\mu_{jk}+\frac{g(C^{*})-g(C_{it})}{\sigma_{ij}}   
\end{equation}
Note that under this parametrization, when $C_{it}=C^{*}$ this model is equivalent to the Bivariate model. For simplicity, we adopt throughout the $\log$ function as a transformation $g()$.
\subsection{The case of networks of tests}\label{Sect:rel_net}
In the case where we are interested in the accuracy and relative performance of $K$ different diagnostic tests, for the same condition, and all tests are binary or continuous but evaluated in the same threshold, the simplest thing one could do in order to analyze all tests in one step is extending the Bivariate model to Bivariate meta-regression (BMR). In this case, the model is defined again by equations \eqref{eq:biv1} and \eqref{eq:biv2}, but with the addition of a covariate representing the different tests on the overall probabilities of a positive test result $\mathbf{m}$. We can equivalently express the logit-transformed probabilities of a positive result of the $k$th test, $p_{ijk}$, as,
$$logit(p_{ijk})=m_{kj}+\epsilon_{ijk},\quad k=1,\dots,K \quad i=1,\dots,I$$
$$(\epsilon_{i0k},\epsilon_{i1k})^{T} \sim N(\mathbf{0},\boldsymbol\Sigma)$$
where $\epsilon_{ijk}$ are random effect terms representing the interaction between study and test. It is also worth noting that this model assumes that all tests have the same covariance matrix $\boldsymbol\Sigma$ which in this case is also of a $2 \times 2$ dimensionality. \par 
This is the current advise according to the Cochrane handbook for systematic reviews of DTA \cite{Macaskill22}. Several other models have been proposed for meta-analyzing multiple tests \citep{lian2018bayesian,nyaga2018anova,menten2015general,ma2018bayesian}, but as mentioned earlier they have not been systematically studied or compared. Here we will further study and extend the ANOVA NMA-DTA model by \citet{nyaga2018anova} as it naturally fits within our proposed framework. This model is an extension of the BMR model and it breaks down the between and within study variances by the inclusion of a further set of study-level random effects, $\eta_{ij}$, as
$$logit(p_{ijk})=m_{kj}+\eta_{ij}+\epsilon_{ijk},\quad k=1,\dots,K \quad i=1,\dots,I$$
$$(\eta_{i0},\eta_{i1})^{T} \sim N(\mathbf{0},\boldsymbol\Sigma), \quad \epsilon_{ijk}\sim N(0,\sigma_{k}^{2})$$
where $\sigma_{k}^{2}$ is the variance of the $k$th test in the network.
\section{NMA-DTA with multiple thresholds}\label{Sect:framework}
In many cases the diagnostic tests network under study might comprise from different types of tests such as binary and continuous reported in multiple different thresholds per study as described in Sections \ref{Sect:intro} and \ref{Sect:rel_work}. We could of course analyze each test independently using, for example, the Bivariate model for the binary ones and the multiple thresholds for the continuous ones. However, being able to analyze all tests together in a single step has several advantages such as efficiency, increased precision for our absolute and relative estimates, the ability to include more tests in the analyses with fewer observations by allowing them to borrow strength from the rest, and ability to rank tests in terms of sensitivity or specificity (or some combination of the 2, e.g. Youden's index). \par 
In this Section we introduce a model that can simultaneously handle both binary and tests reported in multiple thresholds. We explore alternative versions of the model discussing their characteristics and discuss their advantages and disadvantages.
\subsection{Independent meta-analyses in one step (Model 1)}\label{Sect:model_ind}
We begin by considering a model for analyzing all tests in one step that is equivalent to fitting each test independently. Extending the notation of the multiple thresholds model \citep{jones2019quantifying} to multiple tests, the true positive and false positive counts of study $i$, test $k$ for the different thresholds available, $x_{ijkt}$, follow a series of conditional Binomial distributions,
$$x_{ijk1}|p_{ijk1} \sim \mathrm{Binomial}(p_{ijk1},N_{ijk})$$
\begin{equation}\label{eq:cond_binom_net}
    x_{ijkt}|x_{ijk,t-1},p_{ijkt}, p_{ijk,t-1}\sim \mathrm{Binomial}\left(\frac{p_{ijkt}}{p_{ijk,t-1}},x_{ijk,t-1}\right),\quad t=2,\dots,T_{ijk}
\end{equation}
where $p_{ijkt}$ in this case is the probability of a positive test result in study $i$, disease state $j$, test $k$ and threshold $t$. If a test is reported at a single threshold (i.e $T_{ijk}=1$) or is binary, then equation \eqref{eq:cond_binom_net} reduces to a standard Binomial distribution.\par 
Extending equation \eqref{eq:cent} to multiple tests, both continuous and binary, we can express the logit transformed probabilities of a positive test result as,
\begin{equation}\label{eq:logit_ind}
   logit(p_{ijkt})= m_{kj}+\epsilon_{ijk}+\frac{z_{ik}}{\sigma_{ijk}}\log\left( \frac{C_{k}^{*}}{C_{ikt}}\right),\quad \log(\sigma_{ijk})=s_{jk}+u_{ijk} 
\end{equation}

where $m_{jk}$ and $s_{jk}$ are fixed effects terms representing the location and scale parameter of the underlying result of test $k$ in disease status $j$. $C_{ijkt}$ is the $t$th reported threshold in study $i$, test $k$ and disease group $j$, and $C_{k}^{*}$ is the most frequently reported threshold for that test. $z_{ik}$ is a dummy variable that takes the value 1 if an observation corresponds to a continuous test and zero otherwise. Finally, $\epsilon_{ijk}$ and $u_{ijk}$ are random-effects terms representing the interaction between study and test,
$$(\epsilon_{ik1},\epsilon_{ik2}, u_{ik1}, u_{ik2})^{T}\sim N(\mathbf{0},\boldsymbol\Sigma_{k}), $$
Here, $\boldsymbol\Sigma_{k}$ is a $4\times 4$ covariance matrix. In this model there are no common parameters or correlations between tests and it is thus equivalent to fitting a separate model to each of them. If the $k$th test is continuous then is model is equivalent to the multiple thresholds model defined by \eqref{eq:cond_binom_net}, \eqref{eq:cent}.Under this setting, the model implied for the continuous underlying test result of the $l$th patient $y_{ijkl}$, is, $$y_{ijkl} \sim Logistic (\mu_{ijk}*\sigma_{ijk}+\log(C_{k}^{*}),\sigma_{ijk}),\quad \mu_{ijk}=m_{kj}+\epsilon_{ijk}$$
Accordingly, if the $k$th test is binary, then this approach is equivalent to fitting the Bivariate model defined by equations \eqref{eq:biv1}, \eqref{eq:biv2}.\par 
The advantage of this approach is that we can now easily calculate ranking statistics, as seen for example in \citet{owen2018network} or other measures of relative test performance such as differences in sensitivity and specificity, diagnostic odds ratios etc.

\subsection{Along the lines of meta-regression (Model 2)}
Having introduced the unified framework for both binary and continuous tests in the previous section, is very straightforward to adjust it based on the parameterizations of models for multiple tests as described in Section \ref{Sect:rel_net}. An obvious next step would be to consider a model along the lines of meta-regression. Under this setting, the model is defined as in Section \ref{Sect:model_ind} by equations \eqref{eq:cond_binom_net} and \eqref{eq:logit_ind}, but the variance of the random effects however is now shared across all tests as seen below,
$$(\epsilon_{ik1},\epsilon_{ik2},u_{ik1},u_{ik2})^{T}\sim N(\mathbf{0},\boldsymbol\Sigma)$$

This approach allows the inclusion of tests reported in as few as one study in the analysis and will lead to estimates with increased precision. However ,the assumptions that the same variance is shared among all tests can be quite strong, and not always representative of the data \cite{nyaga2018anova}. Note that if all tests are binary and for the case when the threshold of continuous tests is equal to $C^{*}$ (the most frequently reported threshold) this model reduces to the BMR model.

\subsection{ANOVA type parametrization (Model 3)}\label{Sect:m3}
This framework can be naturally extended into an NMA-DTA ANOVA type model following parametrization as seen in \citet{nyaga2018anova}, where some of the variability is attributed to within study factors. In this setting, the model at the first level is defined as in the previous cases by equations \eqref{eq:cond_binom_net}, and the logit-transformed probabilities $p_{ijkt}$ are extended to include additional study-level random effects terms as,
\begin{equation}\label{eq:model3}
    logit(p_{ijkt})= m_{kj}+\epsilon_{ijk}+\eta_{ij}+\frac{z_{ik}}{\sigma_{ikj}}\log\left( \frac{C_{k}^{*}}{C_{ikt}}\right),\quad log(\sigma_{ikj})=s_{jk}+u_{ikj} +\gamma_{ij} 
\end{equation}

$$   ( \eta_{1i}, \eta_{2i}, \gamma_{1i}, \gamma_{2i})^{T} \sim N\left(\boldsymbol0,\mathbf{\Sigma} \right) $$
$$\epsilon_{jik} \sim N(0, \tau m_{jk}^{2}),\quad u_{ijk} \sim N(0,\tau s_{jk}^{2})$$
where $\eta_{ij}$ and $\gamma_{ij}$ are representing the study random effects for the location and scale parameters of the underlying continuous test results distributions. These are assumed to be correlated with each other and between disease states. $\boldsymbol\Sigma$ is a $4\times4 $ variance-covariance matrix shared across tests. $\tau m_{jk}^{2}$ and $\tau s_{jk}^{2}$ are the test-specific variances of the interaction random effect terms for the location and scale parameter respectively. This approach can allow for different variances for each test without falling back to the independent case.\par 
By including the study effect terms we are essentially assuming that the accuracy of tests within the same study is more similar compared to that of tests evaluated in different studies. This model is essentially assuming a linear relationship between the logit transformed probabilities and the log transformed thresholds. From this point of view, this is a random intercept and random slope model with the inclusion of the study effects implying that intercepts and slopes of tests within the same study are more similar to each other than for tests evaluated at different studies. This is more intuitive for the intercept but might be sensible to assume for the slope too if tests have been evaluated at the same individuals since we would expect in this case the spread of disease to be standard within each person. This could also be a sensible assumption even if a study evaluated different tests with different individuals since other setting similarities within each study, such as point of care (primary or secondary) could cause slopes to be more similar within each study.\par
Under this parametrization this model is equivalent to the standard ANOVA model \cite{nyaga2018anova} if all tests are binary and when the threshold is equal to $C_{k}^{*}$, the most frequently reported threshold in this case.\par

\section{Technical details}\label{Sect:tech}
\subsection{Inclusion of tests reported in only one study}
A restriction of the ANOVA type parameterization (Model 3), compared to the meta-regression approach, is that tests evaluated in only one study cannot be included in the analysis. This holds also for the standard ANOVA model \citep{nyaga2018anova}. As explained in section \ref{Sect:m3}, this approach allows the estimation of a separate variance for each test which leads to a total of two test-specific parameters, and thus a minimum of 2 studies per test is required in order to obtain enough degrees of freedom. We can extend this model further so that tests evaluated in only one study can be included, by introducing an extra hierarchical level to this model, by assuming that all tests' variances follow a common distribution (Model 4), 
$$log(\tau m_{jk}) \sim N(m_{aj}, \sigma_{aj}^{2})$$
where $m_{aj}$ and $\sigma_{aj}^{2}$ are the mean and variance of the log-transformed test specific standard deviations $\tau m_{jk}$. Note that for the continuous tests these correspond to the variances of the location parameter of the underlying continuous tests results. It also worth pointing out that this approach can provide sensible results when the number of tests in the network is relatively large.\par 

\subsection{DTA network form}
Each of the models described in the previous sections differs in the assumptions they make about the data, as well as in the type of data they can handle. Details on the assumptions each model makes have been described in the previous sections. For example, the meta-regression approach (Model 2) is the only model that assumes that all tests have the same variance, and the ANOVA approach (Model 3) assumes that tests within a study are more similar compared to tests evaluated in different studies. Since each model estimates a different set of parameters, the form of the DTA network that can be analyzed in each case changes. The ANOVA models are the only approaches described here that require a connected network of tests. Because of the inclusion of the study random effects terms, the model becomes unidentifiable if disconnected tests are included in the analysis. Table \ref{tab:characteristics} summarizes some of the different characteristics each of the models discussed here possesses.  

\begin{table}[!h]
\rowcolors{2}{gray!15}{}
\centering
\caption{\label{tab:characteristics} Summary of models' assumptions and data requirements.}
\resizebox{\textwidth}{!}{%
\begin{tabular}{lrrr}

\textbf{} &\textbf{Test specific variances*} &\textbf{Connected network**}&\textbf{One study tests***}\\
\hline
\textbf{Independent MA} (Model 1) &$\checkmark$ & & \\
\textbf{Meta-regression} (Model 2) & & & $\checkmark$\\
\textbf{ANOVA parametrization}\textsuperscript{\textdagger} (Model 3) &$\checkmark$ & $\checkmark$& \\
\textbf{ANOVA plus parametrization} (Model 4) &$\checkmark$ & $\checkmark$& $\checkmark$\\
\hline


\end{tabular}\hfill
}
{\raggedright \begin{small}\textit{* The model allows for a different variance for each test, ** The model requires a connected network of tests}\\ \textit{*** The model allows the inclusion of tests that have been reported only in one study}\\ \textsuperscript{\textdagger}\textit{Including studies reporting accuracy of a single test results in an identifiable model as long as the test is part of the connected network.}\\  \end{small}\par}
\end{table}

\subsection{Different covariance structures}
All models described here are presented in their most general form. These models require the estimation of a relatively large amount of parameters which might not always be suitable or feasible given the available data. This might be especially the case for the parts of the model capturing the relationship of test accuracy and threshold for continuous tests. In general, in all cases presented here, the models can be simplified by considering structured or restricted forms of the $4\times 4$ variance-covariance matrices $\Sigma$ or $\Sigma_{k}$ as shown in \citet{jones2019quantifying} and \citet{derezea2024technical}. Empirical applications of the \citet{jones2019quantifying} model for multiple thresholds show that in many cases the dependencies between the scale and location parameters could be dropped.\par 
The ANOVA parametrization approach (Model 3) could be further simplified by dropping some of the random effects terms, as it might not always be necessary to have the same sets of fixed and random effects on both $\sigma_{ij}$ and $\mu_{ij}$. We could for example remove the study random effects from the standard deviation. Based on these, equation \eqref{eq:model3} simplifies to, 
$$logit(p_{ijkt})= m_{kj}+\epsilon_{ijk}+\eta_{ij}+\frac{z_{ik}}{\sigma_{ikj}}\log\left( \frac{C_{k}^{*}}{C_{ikt}}\right),\quad log(\sigma_{ikj})=s_{jk}+u_{ikj}  $$
$$(\eta_{i1},\eta_{i2})^{T} \sim N(\mathbf{0},\boldsymbol\Sigma)$$
which requires the estimation of a variance-covariance matrix of only a $2\times 2$ dimensionality.
\subsection{Choice of hyper-priors}
Throughout this paper, we will be working within a fully Bayesian framework. To complete the model, we assign the following sets of hyper-priors.
$$m_{kj}\sim N(0,1000), \quad s_{jk}\sim N(0,1000)$$
$$\boldsymbol\Sigma \sim Wishart (R,\nu)$$
For the general version of model 3, $R$ could be a $4\times 4$ identity matrix, and the degrees of freedom $\nu$ set equal to 4.
$$\tau s_{jk}\sim U(0,5)$$
$$\exp (m_{aj}) \sim Uniform(0,5), \quad \sigma_{aj} \sim Uniform (0,20)$$
These hyper-priors have been mainly chosen for computational convenience. The width of the priors of the standard deviation parameters was chosen based on guidance from the Technical Support document 25 \cite{derezea2024technical}.  

\section{Application to real data}\label{Sect:applic}
\subsection{Case study 1: Hepatocellular carcinoma (HCC) tests network}
\begin{figure}[h!]
    \centering
    \includesvg[width=300pt]{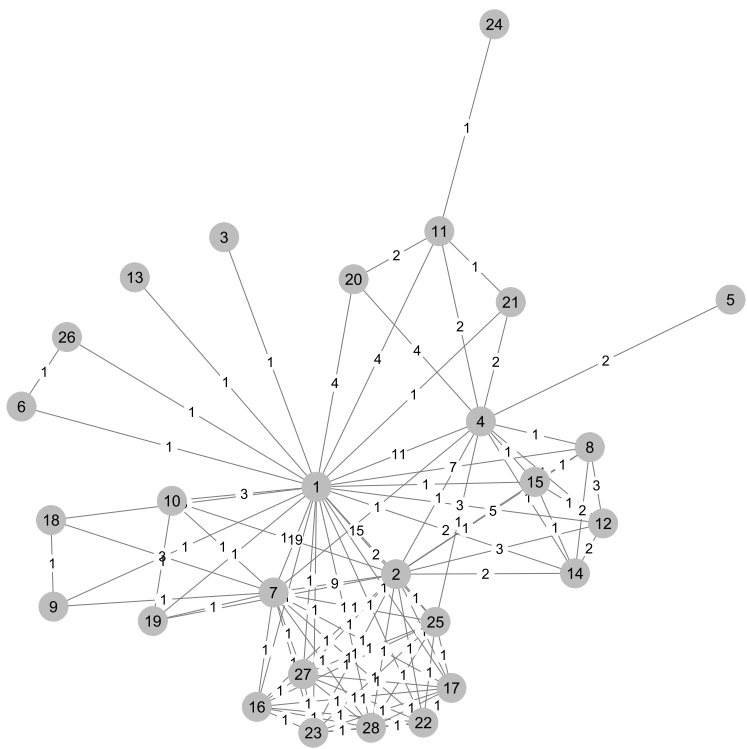}

    \caption{Network from the HCC review: \textbf{(1)} AFP, \textbf{(2)} AFP-L3, \textbf{(3)} AFP progression rate (ng/ml per month), \textbf{(4)} B-mode US, \textbf{(5)} CEUS, \textbf{(6)} Combined AFP index (serial AFP and AFP), \textbf{(7)} DCP, \textbf{(8)} DCP (ng/ml), \textbf{(9)} Model based on DCP, AFP, gender, and age, \textbf{(10)} Doylestown algorithm, \textbf{(11)} Dynamic contrast-enhanced MRI, \textbf{(12)} GALAD, \textbf{(13)} HCC-ART, \textbf{(14)} HES algorithm, \textbf{(15)} Longitudinal GALAD, \textbf{(16)} mFB-I, \textbf{(17)} mFB-J, \textbf{(18)} Model based on AFP and DCP, \textbf{(19)} Model based on age, gender, AFP and DCP, \textbf{(20)} Multiphase HCC-specific protocol CT, \textbf{(21)} Noncontrast MRI, \textbf{(22)} PEB algorithm (AFP), \textbf{(23)} PEB algorithm (DCP), \textbf{(24)} PM-DL model, \textbf{(25)} Serial AFP, \textbf{(26)} Serial AFP – any increase, \textbf{(27)} uFB (AFP), \textbf{(28)} uFB (DCP)}

    \label{fig:net-hcc}
\end{figure}

In a recent systematic review of diagnostic tests for detecting hepatocellular carcinoma (HCC) in patients with cirrhosis \cite{rogers2025cirrhosis}, there were 97 tests evaluated in 160 studies forming a connected evidence synthesis network. For demonstration purposes here we use a restricted dataset comprising only of the continuous conventional biomarkers, imaging and algorithm-based tests. This lead to a network of 28 tests evaluated in 130 studies, visualized in Figure \ref{fig:net-hcc}. Four of these tests were continuous biomarkers with have their accuracy reported in multiple thresholds per study. Test alpha-fetoprotein (AFP) in particular was reported in a total of 157 different thresholds with the maximum number of thresholds reported in a single study being 48. This was also the most commonly reported test in the review.\par 

We fitted the three different models of our proposed framework: Independence (Model 1), Meta-regression (Model 2) and ANOVA plus parametrization (Model 4), along with some more simplified versions of the latter. We compared these models based on the goodness of fit using the DIC criterion, where lower values are an indication of a better fit (after penalizing for complexity). A simplified version of model 4 with a restricted covariance matrix was selected based on DIC. The detailed results can be found in the appendix. \par
\begin{figure}[h!]
    \centering
    \includesvg[width=510pt]{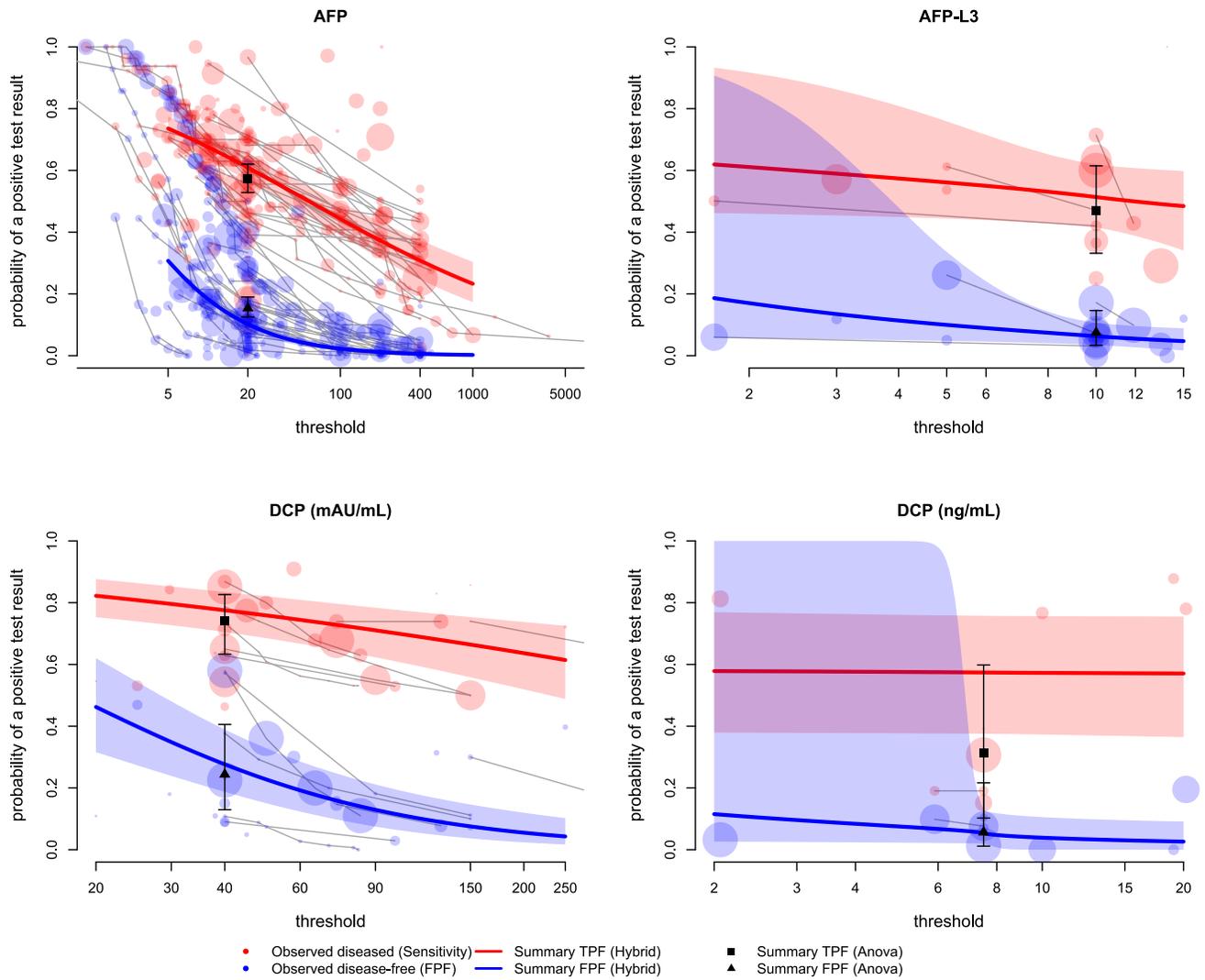}

    \caption{Results for the continuous tests of the hcc review }

    \label{fig:cont-res-hcc}
\end{figure}

\begin{figure}[h!]
    \centering
    \includesvg[width=510pt]{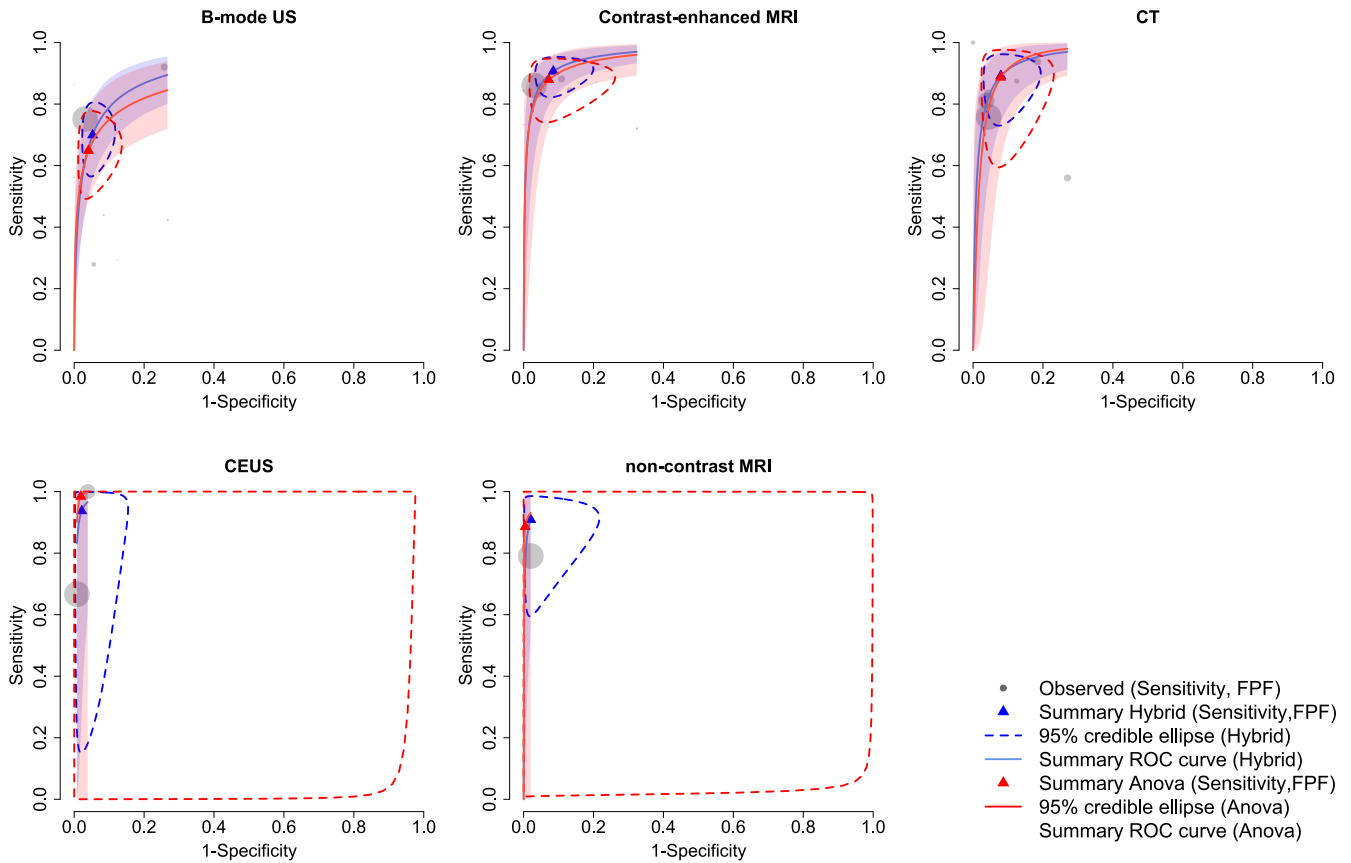}

    \caption{Results for the imaging tests of the hcc review }

    \label{fig:im-res-hcc}
\end{figure}

 
The results for continuous tests can be seen in Figure \ref{fig:cont-res-hcc}, where the pooled sensitivity and FPF is depicted across the whole range of observed thresholds along with their corresponding 95\% credible intervals (CrIs). For the five imaging tests, we see in Figure \ref{fig:im-res-hcc} pooled estimates of sensitivity and specificity in the receiver operating characteristic (ROC) space along with 95\% credible ellipses. Their corresponding  summary ROC curve is also shown along with its corresponding 95\% CrIs which were calculated according to the TSD25 guidance \cite{derezea2024technical}. The pooled estimates of sensitivity and specificity for the remaining tests can be found in the appendix. \par 

For a comparison, we further fitted the standard approach for binary tests by selecting only the most commonly reported thresholds for the continuous tests using the standard \citet{nyaga2018anova} ANOVA model. This approach resulted in a large amount of evidence effectively being discarded as only half of the tests were forming a connected network in this case. See Figures \ref{fig:cont-res-hcc} and \ref{fig:im-res-hcc} for a comparison between the two approaches with our hybrid method returning almost all of the time estimates of higher precision as reflected by the tighter credible intervals. 

\subsection{Case study 2: Prostate cancer network}
\begin{figure}[h!]
    \centering
    \includesvg[width=250pt]{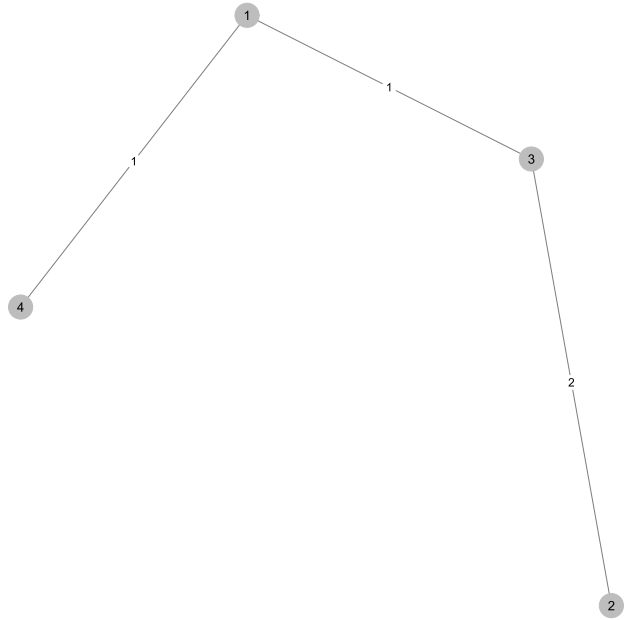}

    \caption{Network from the prostate review: \textbf{(1)} 4K, \textbf{(2)} PCA3, \textbf{(3)} PHI, \textbf{(4)} SelectMDx}

    \label{fig:net-prost}
\end{figure}

The second case study was based on a systematic review by \citet{kawada2024diagnostic} on tests for detecting prostate cancer. This review consisted of continuous biomarkers only, four of which form a connected evidence synthesis network as seen in Figure \ref{fig:net-prost}. This is a much sparser network compared to that of the first case study, consisting of 37 studies. Although all four tests were continuous biomarkers, we treated test SelectMDx as binary because its accuracy was reported in a very small number of thresholds based on which we could not reliably estimate a relationship between the probability of a positive test result and threshold.\par 

\begin{figure}[h!]
    \centering
    \includesvg[width=400pt]{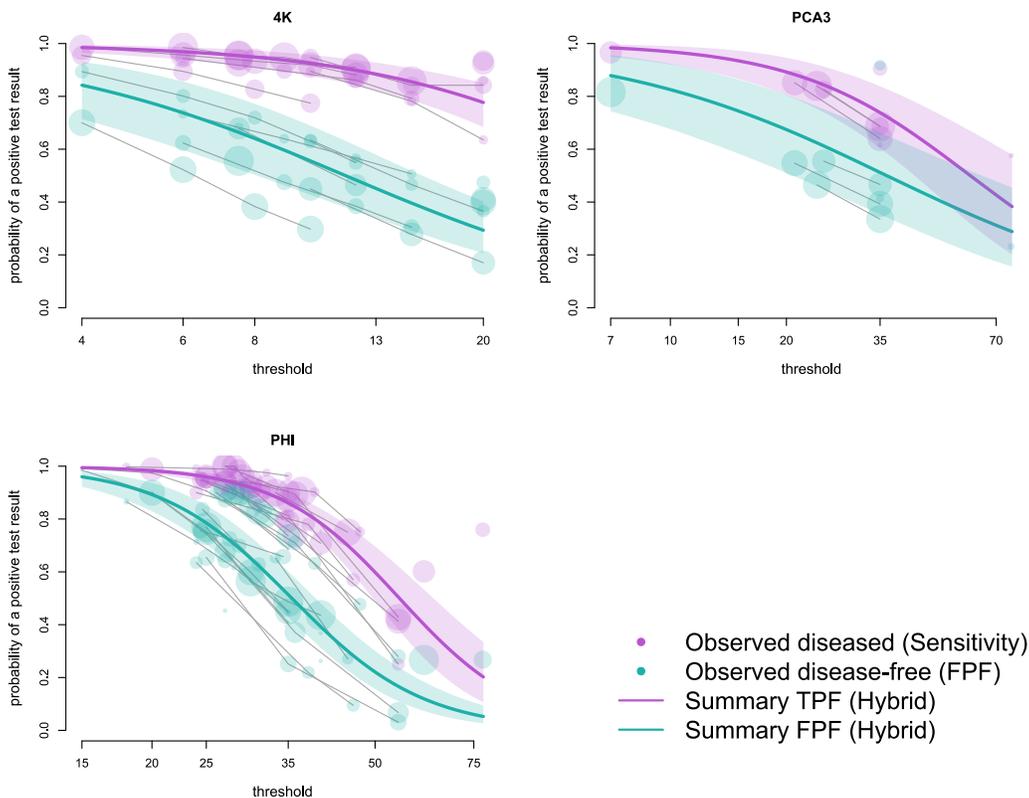}

    \caption{Results for the continuous tests of the prostate cancer review }

    \label{fig:cont-res-prost}
\end{figure}

We fitted the same models as in the previous case. The meta-regression parametrization returned the lowest DIC value in this case which is not surprising given the sparsity of the dataset. More details can be found in the appendix. The results for the three multiple threshold tests can be seen in Figure \ref{fig:cont-res-prost}.\par 
In this case we could not compared results to the standard \citet{nyaga2018anova} ANOVA model. The reason for this was that once the most frequently reported threshold from each test was selected, none of the tests formed a connected evidence synthesis network. Pooled estimates of sensitivity and specificity for all methods fitted for this dataset can be found in the appendix.



\section*{Acknowledgments}
HEJ and ED were supported by an MRC-NIHR New Investigator Research Grant (MR/T044594/1). This work was additionally supported by the NIHR Health Technology Assessment programme (project reference NIHR134670). 

\section*{Data availability}
The Data and R/jags code to reproduce some of the analyses in this paper are available on Git-hub at:  \href{https://github.com/FeniaDerezea/NMA-DTA_mult_thres}{github.com/FeniaDerezea}
\newpage 

\newpage
\bibliography{example}
\newpage
\appendix
\section{Appendix}

\subsection{Example 1: Model comparison additional results}

\begin{table}[!h]
\rowcolors{2}{gray!15}{}
\centering
\caption{\label{tab:gof} Goodness of fit comparison for the HCC dataset.}
\begin{tabular}{lrrrrr}

\textbf{}&\textbf{V1}&\textbf{V2} &\textbf{V3} &\textbf{Met-reg}\\
\hline
\textbf{Residual Deviance}&1071.7  &1071.7 &1084.5 &1081.5  \\
\textbf{pV}& 570.0 &571.6 &593.5 &579.2  \\
\textbf{DIC*}& 1641.7 &1643.3 &1678.0 &1660.7  \\
\hline


\end{tabular}\hfill
\end{table}

\begin{table}[hbp]
\rowcolors{2}{gray!15}{}
\centering
\caption{\label{tab:tp_comparison} Sensitivity and Specificity point estimates along with 95\% CrIs for the HCC network.}
\resizebox{\textwidth}{!}{%
\begin{tabular}{lrrrrrrrrr}
\multicolumn{2}{c}{} &
      \multicolumn{2}{c}{Model V2}&\multicolumn{2}{c}{Met-reg}&\multicolumn{2}{c}{Ind}&\multicolumn{2}{c}{Nyaga Anova}\\  
\textbf{Test}&\textbf{Studies}  &\textbf{Sens}&\textbf{Spec} &\textbf{Sens} &\textbf{Spec}&\textbf{Sens}&\textbf{Spec}&\textbf{Sens}&\textbf{Spec}\\
\hline

\textbf{AFP at 20ng/mL}&  & 0.61 (0.57,0.65)& 0.90 (0.87,0.92)&   0.61 (0.57,0.65)& 0.90 (0.87,0.92)& 0.61 (0.57,0.65)& 0.89 (0.87,0.92) & 0.57 (0.53,0.62)& 0.85 (0.81,0.87)\\
\textbf{AFP-L3 at 10\%}&  &0.51 (0.41,0.62)& 0.94 (0.89,0.97)&  0.46 (0.34,0.59)& 0.94 (0.90,0.97)& 0.46 (0.37,0.57)& 0.93 (0.90,0.96) &0.47 (0.33,0.61)& 0.93 (0.85,0.97)\\
\textbf{AFP progression rate}&  &0.64 (0.11,0.96)& 1.00 (0.97,1.00) & 0.67 (0.26,0.92)&1.00 (0.97,1.00)& -& - & -& -\\
\textbf{B-mode US}&  & 0.70 (0.59,0.79)& 0.95 (0.90,0.97) &  0.62 (0.50,0.72)&0.96 (0.93,0.98)&0.62 (0.49,0.73)& 0.96 (0.92,0.98) & 0.65 (0.54,0.75)&0.96 (0.91,0.98)\\
\textbf{CEUS}&  & 0.94 (0.52,1.00)& 0.98 (0.89,1.00)&  0.91 (0.66,0.98) &0.98 (0.91,1.00)& -& - & 0.99 (0.01,1.00)&0.98 (0.05,1.00)\\
\textbf{Combined AFP index}&  & 0.81 (0.27,0.98)& 0.62 (0.15,0.94)& 0.80 (0.39,0.96)&0.62 (0.14,0.94)& -& - & -& -\\
\textbf{DCP at 40mAU/mL}&   & 0.78 (0.70,0.83)& 0.72 (0.61,0.81)&0.75 (0.67,0.82)&0.75 (0.64,0.83)&0.74 (0.67,0.81)&0.75 (0.64,0.84) & 0.74 (0.63,0.83)&0.76 (0.59,0.87)\\
\textbf{DCP at 7.5ng/mL}&  & 0.57 (0.38,0.76)& 0.95 (0.87,0.98)& 0.54 (0.38,0.70)&0.95 (0.88,0.98)&0.54 (0.27,0.78)&0.94 (0.83,0.99) & 0.31 (0.10,0.60)&0.94 (0.78,0.99)\\
\textbf{Model1}&  & 0.86 (0.34,0.98)& 0.90 (0.47,0.99)&0.89 (0.56,0.98)&0.93 (0.56,0.99)& -& - & -& -\\
\textbf{Doylestown}&  & 0.50 (0.18,0.71)&0.94 (0.82,0.98) & 0.47 (0.22,0.73)&0.94 (0.82,0.98)&0.44 (0.01,0.80)&0.94 (0.84,0.98) &0.39 (0.00,0.98)&0.93 (0.37,1.00)\\
\textbf{Contrast MRI}&  & 0.91 (0.84,0.95)& 0.92 (0.83,0.96)& 0.86 (0.76,0.92)&0.93 (0.86,0.97)&0.86 (0.79,0.91)&0.93 (0.87,0.97) &0.88 (0.79,0.94)&0.93 (0.84,0.98)\\
\textbf{GALAD}&  & 0.81 (0.64,0.91)& 0.79 (0.53,0.93) &0.68 (0.42,0.86)&0.81 (0.52,0.94)&0.68 (0.39,0.87)&0.81 (0.54,0.93) &0.75 (0.30,0.96)&0.78 (0.09,0.99)\\
\textbf{HCC-ART}&  & 0.85 (0.40,0.98)&0.84 (0.35,0.98)& 0.90 (0.59,0.98)&0.82 (0.30,0.98) & -& - & -& -\\
\textbf{HES }&  & 0.49 (0.20,0.77)&0.91 (0.69,0.98) &0.33 (0.11,0.66)&0.92 (0.70,0.98)& -& - &0.43 (0.00,1.00)&0.90 (0.02,1.00)\\
\textbf{Longitudinal GALAD}&  & 0.70 (0.18,0.96)&0.91 (0.53,0.99)& 0.63 (0.19,0.92)&0.90 (0.48,0.99)& -& - & -& -\\
\textbf{mFB-I}& & 0.72 (0.21,0.96)&0.91 (0.55,0.99)& 0.60 (0.19,0.91)&0.90 (0.49,0.99)& - & - & -& -\\
\textbf{mFB-J}&  & 0.72 (0.16,0.96)&0.91 (0.54,0.99)&0.61 (0.21,0.90)&0.90 (0.46,0.99) &-&-& -& -\\
\textbf{Model2}&  & 0.75 (0.56,0.88)& 0.86 (0.65,0.96)& 0.74 (0.50,0.89)&0.87 (0.64,0.96) &0.74 (0.44,0.92)&0.87 (0.63,0.97)& 0.76 (0.21,0.98)&0.86 (0.17,1.00)\\
\textbf{Model3}&  & 0.66 (0.14,0.94)& 0.90 (0.54,0.99) &0.68 (0.26,0.93)&0.89 (0.48,0.99) &-&-& -& -\\
\textbf{CT}&  & 0.89 (0.79,0.96)& 0.92 (0.84,0.96) & 0.87 (0.76,0.93)&0.92 (0.82,0.96)&0.87 (0.72,0.96)&0.92 (0.83,0.97) &0.89 (0.72,0.97)& 0.92 (0.83,0.97)\\
\textbf{Noncontrast MRI}&  & 0.91 (0.68,0.98)& 0.98 (0.89,1.00)& 0.82 (0.53,0.95)&0.99 (0.93,1.00)& -& - &0.88 (0.01,1.00)&1.00 (0.06,1.00)\\
\textbf{PEB (AFP)}& & 0.85 (0.32,0.98)& 0.91 (0.54,0.99) &0.78 (0.34,0.96)&0.90 (0.47,0.99)&-& - & -& -\\
\textbf{PEB (DCP)}&  & 0.72 (0.17,0.96)&0.91 (0.52,0.99) & 0.62 (0.21,0.91)& 0.90 (0.47,0.99)& -& - & -& -\\
\textbf{PM-DL model}&  & 0.96 (0.60,1.00)& 1.00 (0.90,1.00)& 0.95 (0.67,1.00)&1.00 (0.94,1.00)&-&- & -& -\\
\textbf{serial AFP 1}&  & 0.86 (0.12,1.00)& 0.90 (0.64,0.98) & 0.87 (0.62,0.96)&0.86 (0.55,0.97)& -& - & 0.84 (0.00,1.00)&0.90 (0.01,1.00)\\
\textbf{Serial AFP 2}& & 0.69 (0.17,0.96)& 0.70 (0.20,0.96)& 0.68 (0.25,0.93)&0.71 (0.20,0.96)& -& - & -& -\\
\textbf{uFB (AFP)}& & 0.88 (0.36,0.99)& 0.91 (0.55,0.99)& 0.82 (0.41,0.97)&0.89 (0.47,0.99)&-&- & -& -\\
\textbf{uFB (DCP)}&  & 0.81 (0.23,0.98)& 0.91 (0.53,0.99)& 0.71 (0.27,0.94)& 0.90 (0.50,0.99)&-&- & -& -
\\
\hline

\end{tabular}\hfill
}
\end{table}

\subsection{Example 2: Model comparison additional results}
\begin{table}[!h]
\rowcolors{2}{gray!15}{}
\centering
\caption{\label{tab:gof2} Goodness of fit comparison for the prostate cancer dataset.}
\begin{tabular}{lrrrrr}

\textbf{}&\textbf{V1}&\textbf{V2} &\textbf{V3} &\textbf{Met-reg}\\
\hline
\textbf{Residual Deviance}&211.1  &211.0 &210.8 & 208.7 \\
\textbf{pV}& 118.2 &118.8 &115.9 &113.0  \\
\textbf{DIC*}& 329.3 &329.8 &326.7 & 321.7 \\
\hline


\end{tabular}\hfill
\end{table}

\begin{table}[hbp]
\rowcolors{2}{gray!15}{}
\centering
\caption{\label{tab:tp_comparison2} Sensitivity and Specificity point estimates along with 95\% CrIs for the prostate cancer network.}
\resizebox{\textwidth}{!}{%
\begin{tabular}{lrrrrrrrrr}
\multicolumn{2}{c}{} &
      \multicolumn{2}{c}{Model V3}&\multicolumn{2}{c}{Met-reg}&\multicolumn{2}{c}{Ind}&\multicolumn{2}{c}{Nyaga Anova}\\  
\textbf{Test}&\textbf{Studies}  &\textbf{Sens}&\textbf{Spec} &\textbf{Sens} &\textbf{Spec}&\textbf{Sens}&\textbf{Spec}&\textbf{Sens}&\textbf{Spec}\\
\hline

\textbf{4K at 20}&  &0.70 (0.57,0.82) & 0.77 (0.68,0.83)&  0.78  (0.68,0.85)& 0.71 (0.60,0.79)& 0.78 (0.67,0.87)& 0.70 (0.60,0.80) &  -&  -\\
\textbf{PCA3 at 35}&  &0.80 (0.67,0.88)& 0.41 (0.20,0.67)& 0.74  (0.60,0.84)&  0.49 (0.34,0.65)& 0.73 (0.57,0.85)& 0.48 (0.26,0.73) & -&  -\\
\textbf{PHI at 35}&  &0.88 (0.83,0.91)&  0.47 (0.39,0.55) & 0.86 (0.82,0.90)& 0.49 (0.40,0.57)& 0.86 (0.81,0.90)& 0.49 (0.41,0.57) & -& -\\
\textbf{SelectMDx at 7.5}&  & 0.47 (0.01,0.98)& 0.80  (0.01,1.00) &  0.55 (0.19,0.86)& 0.76 (0.35,0.95)& -& - & -& -
\\
\hline

\end{tabular}\hfill
}
\end{table}

\end{document}